\documentclass[preprint,12pt]{elsarticle}

\usepackage{amssymb}
\usepackage{amsthm}
\usepackage{xspace}

\newcommand{\dau}{\mbox{$d$$+$Au}\xspace}
\newcommand{\pt}{\mbox{$p_T$}\xspace}
\newcommand{\auau}{\mbox{Au$+$Au}\xspace}
\newcommand{\cucu}{\mbox{Cu$+$Cu}\xspace}
\newcommand{\rda}{\mbox{$R_{dA}$}\xspace}
\newcommand{\raa}{\mbox{$R_{AA}$}\xspace}
\newcommand{\ncoll}{\mbox{$\langle N_{\rm coll}\rangle$}\xspace}
\newcommand{\npart}{\mbox{$\langle N_{\rm part}\rangle$}\xspace}

\newcommand{\pp}{\mbox{$p$$+$$p$}\xspace}

\journal{Nuclear Physics A}

\begin{document}

\begin{frontmatter}



\title{PHENIX recent heavy flavor results}


\author{Sanghoon Lim for the PHENIX collaboration}

\address{Physics Department, Yonsei University, Seoul 120-749, Korea}

\begin{abstract}
Cold nuclear matter (CNM) effects provide an important baseline for the interpretation of data in heavy ion collisions. Such effects include  nuclear shadowing, Cronin effect, and initial patron energy loss, and it is interesting to study the dependence on impact parameter and kinematic region. Heavy quark production is a good measurement to probe the CNM effects particularly on gluons, since heavy quarks are mainly produced via gluon fusions at RHIC energy. The PHENIX experiment has experiment has ability to study the CNM effects by measuring heavy quark production in \dau collisions at variety of kinematic ranges. Comparisons of heavy quark production at different rapidities allow us to study modification of gluon density function in the Au nucleus depending on momentum fraction. Furthermore, comparisons to the results from heavy ion collisions (\auau and \cucu) measured by PHENIX provide insight into the role of CNM effects in such collisions. Recent PHENIX results on heavy quark production are discussed.

\end{abstract}


\end{frontmatter}


\section{Introduction}
\label{intro}
Heavy quarks, mostly charm and bottom at RHIC energies, are produced via gluon fusion in the early stage of heavy-ion collisions.
Therefore, they are good tools to study the evolution of medium produced in heavy-ion collisions. 
Due to their large masses, heavy quark production is naturally a hard process so that it can be described by perturbative QCD calculations.
These theoretical approaches can be tested by comparison to the measurement of heavy quark production in \pp collisions.
The results in \pp collisions also provide a baseline in order to quantify nuclear effects in other collision systems.
In a hot and dense medium, ``dead cone effect'' predict that bottom quarks will lose less energy than charm quarks ($R_{AA}^{c}<R_{AA}^{b}$) due to a limited range of gluon radiation~\cite{deadcone:2001}.
The precise measurement in heavy-ion collisions will provide essential information about energy loss mechanism of heavy quarks inside the produced medium.
The production of heavy quarks can be modified at final stage due to the hot and dense medium as well as at initial stage before heavy-ion collisions.
In order to distinguish the initial- and final-state modification, \dau collisions are used as a control experiment.

The PHENIX experiment has a suitable design to measure leptons from semi-leptonic decay of open heavy-flavor hadrons. 
In central arms at mid-rapidity ($|\eta|<0.35$) and muon arms at forward rapidity ($1.2<|\eta|<2.2$), heavy-flavor electrons and muons are measured, respectively.
Hadron cocktail method is used for background estimation at both rapidity regions, and a converter method is also used for heavy-flavor electron analysis at mid-rapidity.
For the baseline measurements, PHENIX measured heavy-flavor electrons and muons at mid- and forward rapidity, respectively~\cite{ppg065,ppg117}, and the results of charm cross section are consistent with FONLL calculations within uncertainties.

\section{Heavy-ion results}
\label{heavyion}
\begin{figure}
\centering
\includegraphics[width=0.51\textwidth,clip]{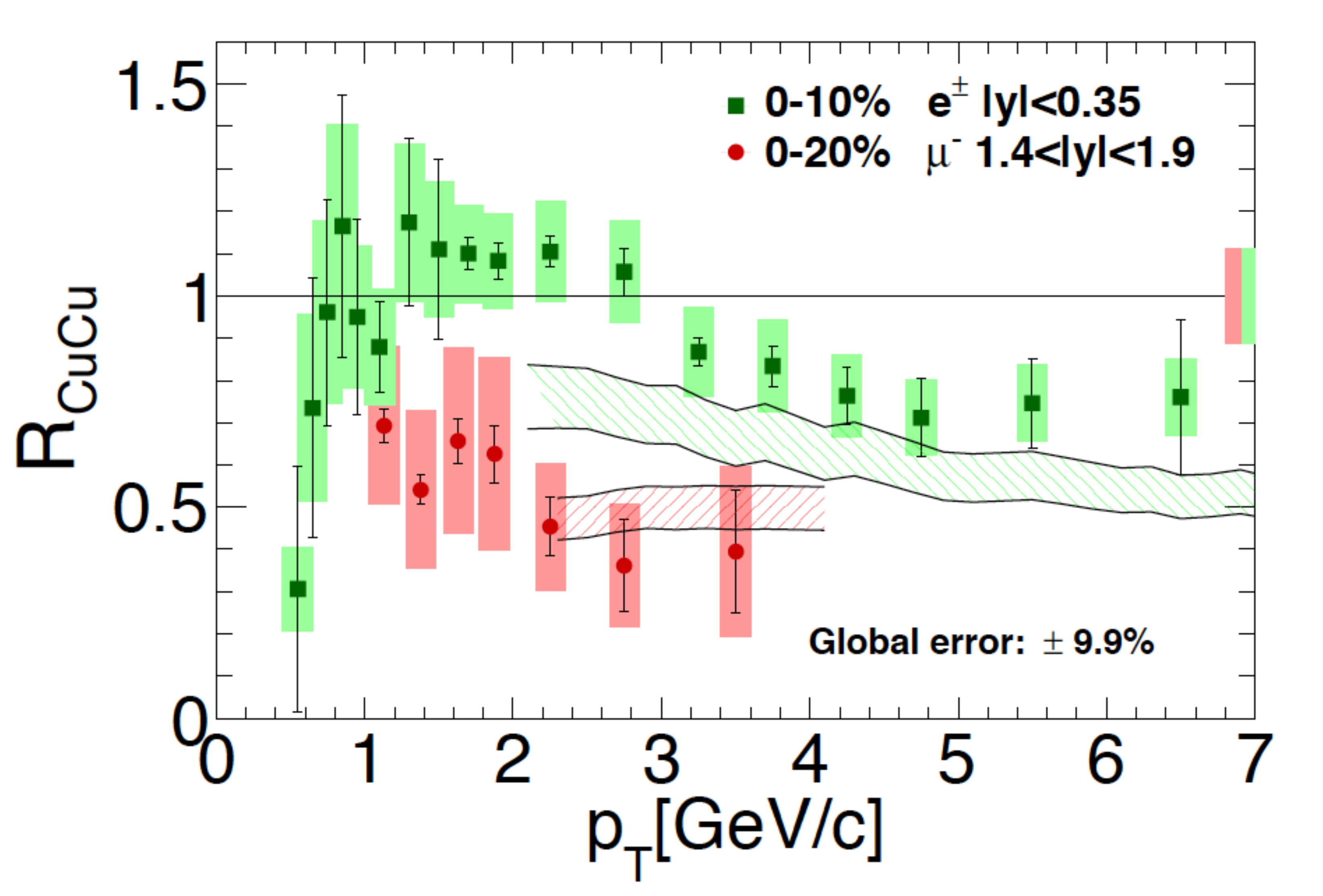}
\includegraphics[width=0.48\textwidth,clip]{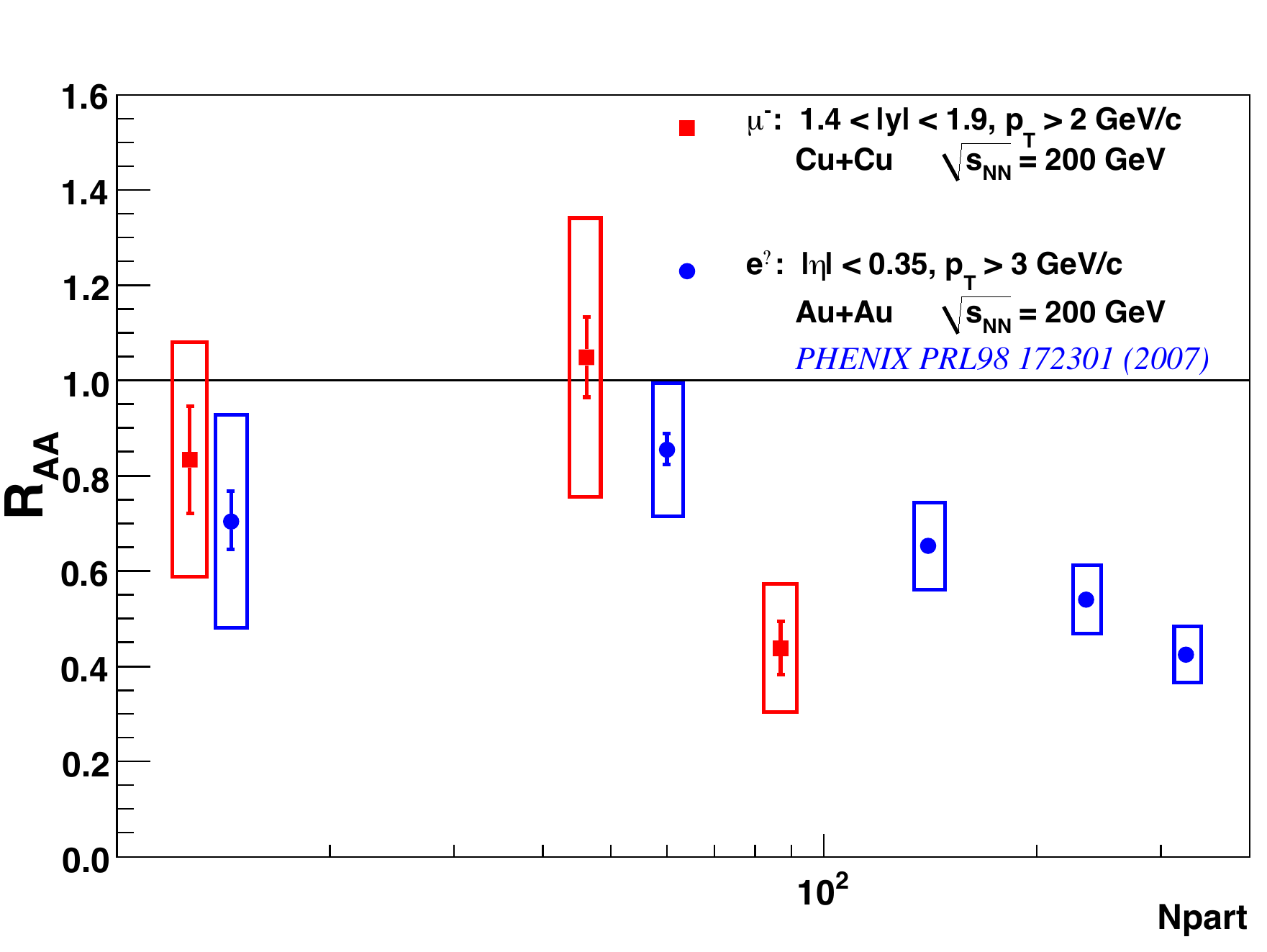}
\caption{Left: \raa of heavy-flavor leptons as a function of \pt in central \cucu collisions at mid- and forward rapidity regions. Right: Comparion of \raa of heavy-flavor leptons as a function of \npart between \auau collisions at mid-rapidity and \cucu collisions at forward rapidity.}
\label{fig1}
\end{figure}

Previously, PHENIX measured heavy-flavor electrons in \auau collisions at mid-rapidity~\cite{ppg066}.
In this measurement, a significant suppression is observed in high \pt region, and the values of \raa at $\pt>5~{\rm GeV/c}$ is almost similar with that for $\pi^{0}$.
Recently, PHENIX has shown heavy-flavor electron measurements in smaller systems, \cucu collisions.
The left panel in Fig.~\ref{fig1} shows \raa of heavy-flavor leptons in central \cucu collisions at mid- (squares) and forward (circles) rapidity regions~\cite{ppg150,ppg117}. 
At mid-rapidity, a small suppression is observed at $\pt>3~{\rm GeV/c}$, where as a huge suppression is seen at forward rapidity.
Two bands in the same figure represent pQCD calculations considering both collisional and radiative energy loss in the hot and dense medium as well as cold nuclear matter (CNM) effects~\cite{vitev:2009}.
The theoretical predictions are qualitively consistent with the data at both rapidity regions.
\begin{figure}
\centering
\includegraphics[width=0.6\textwidth,clip]{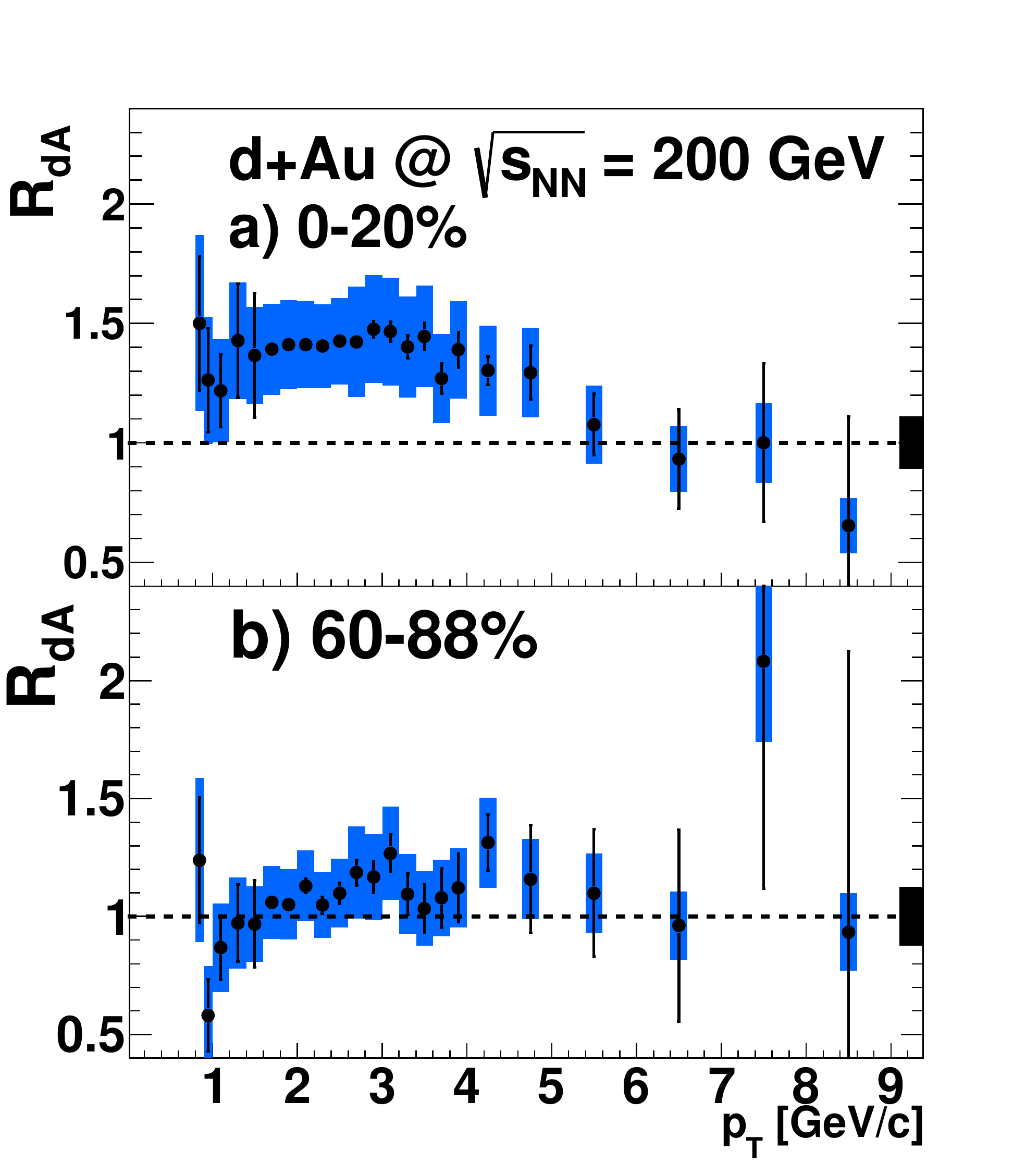}
\caption{\rda as a function of \pt of heavy-flavor electrons at mid-rapidity in the most central (top) and the most peripheral (bottom) centrality classes.}
\label{fig2}
\end{figure}

The right panel in Fig.~\ref{fig1} shows comparison of \raa as a function of \npart between \cucu results at forward and \auau results at mid-rapidity.
As can be seen in this plot, it is interesting that \raa in central \cucu collisions at forward rapidity is comparable with that in central \auau collisions at mid-rapidity.
The additional CNM effects at forward rapidity may contribute the large suppression of heavy-flavor muons.

\section{\dau results}
\label{dau}
\begin{figure}
\centering
\includegraphics[width=1.0\textwidth,clip]{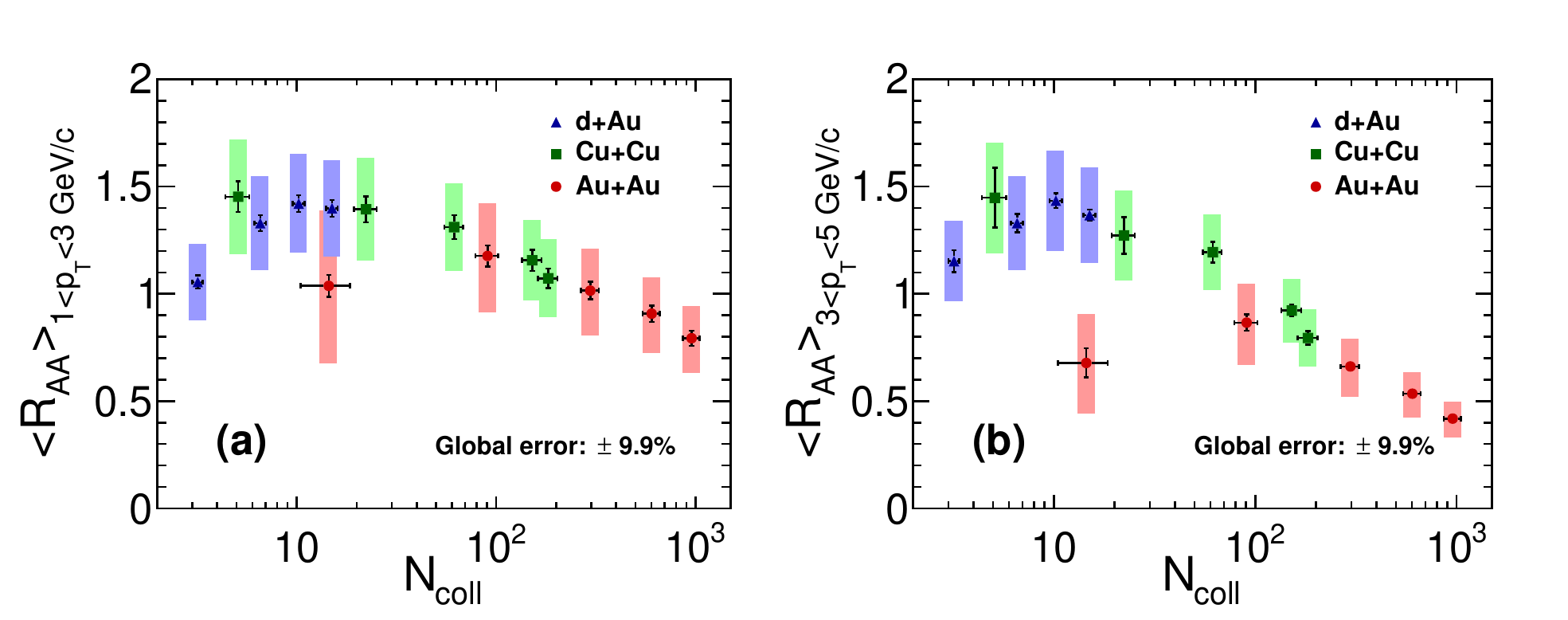}
\caption{\raa of heavy-flavor electrons as a function of \ncoll in various collision systems.}
\label{fig4}
\end{figure}
In order to correctly interpret and better understand the results in heavy-ion collisions, the results in \dau collisions with minimal effects of hot and medium are needed.
During the \dau run in 2008, PHENIX collected large number of event samples, so many interesting results to study CNM effect have been came out.
Figure~\ref{fig2} shows \rda as a function of \pt for heavy-flavor electrons in the most central (top) and most peripheral (bottom) centrality classes measured at mid-rapidity~\cite{ppg131}.
In central \dau collisions, heavy-flavor electron production is clearly enhanced at moderate \pt region than the scaled \pp results, whereas no overall modification is observed in the most peripheral centrality class.
By comparing the results in heavy-ion collisions, one can conclude that the suppression seen in central \auau collisions is due to the hot and dense medium.

Figure~\ref{fig4} shows \raa as a function of \ncoll in \dau, \cucu, and \auau collisions for two different \pt ranges at mid-rapidity. 
Between the enhancement in \dau collisions and the suppression in central \auau collisions, a smooth transition between cold and hot nuclear matter effects is seen in \cucu collisions.

\begin{figure}
\centering
\includegraphics[width=0.49\textwidth,clip]{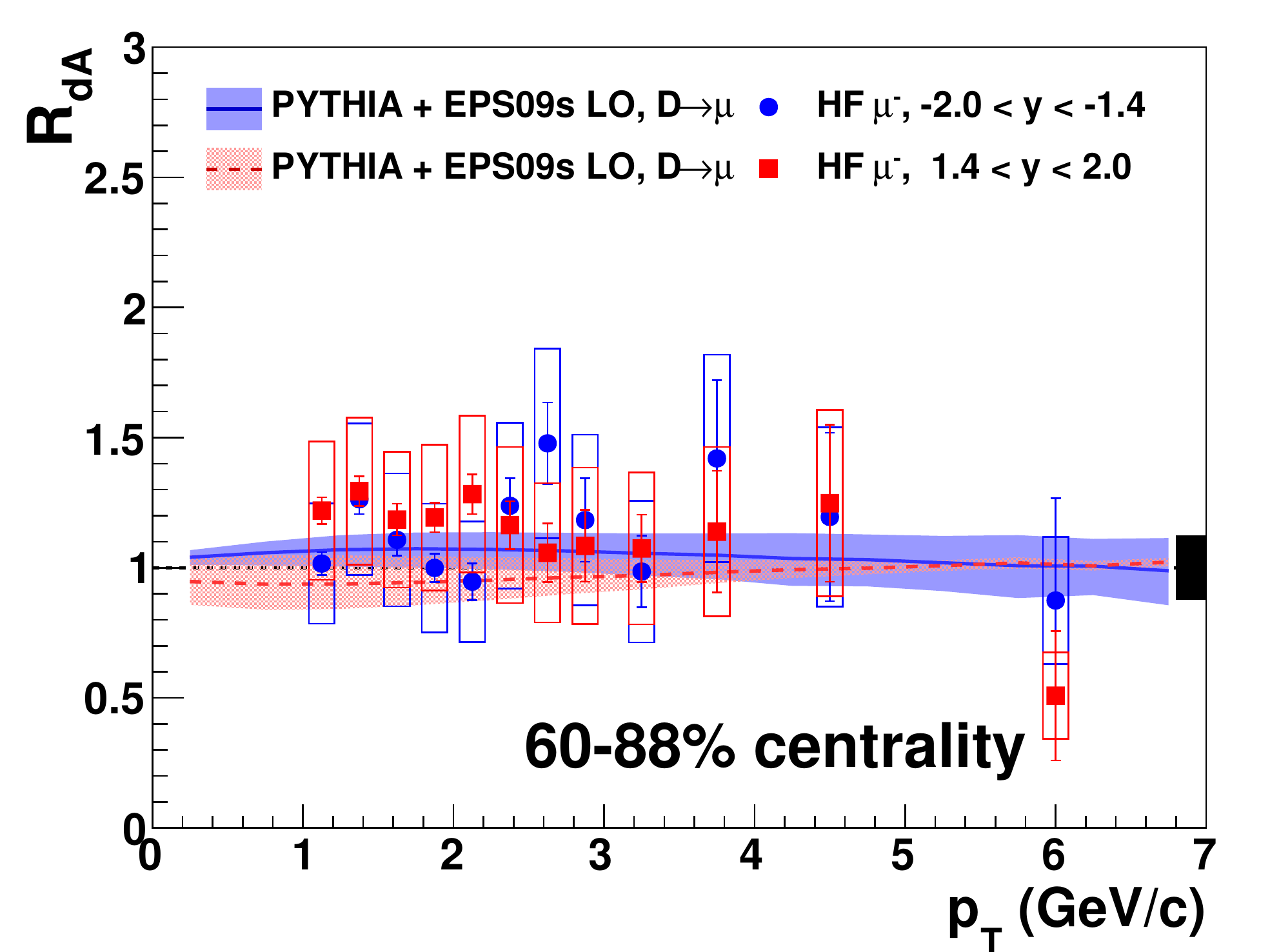}
\includegraphics[width=0.49\textwidth,clip]{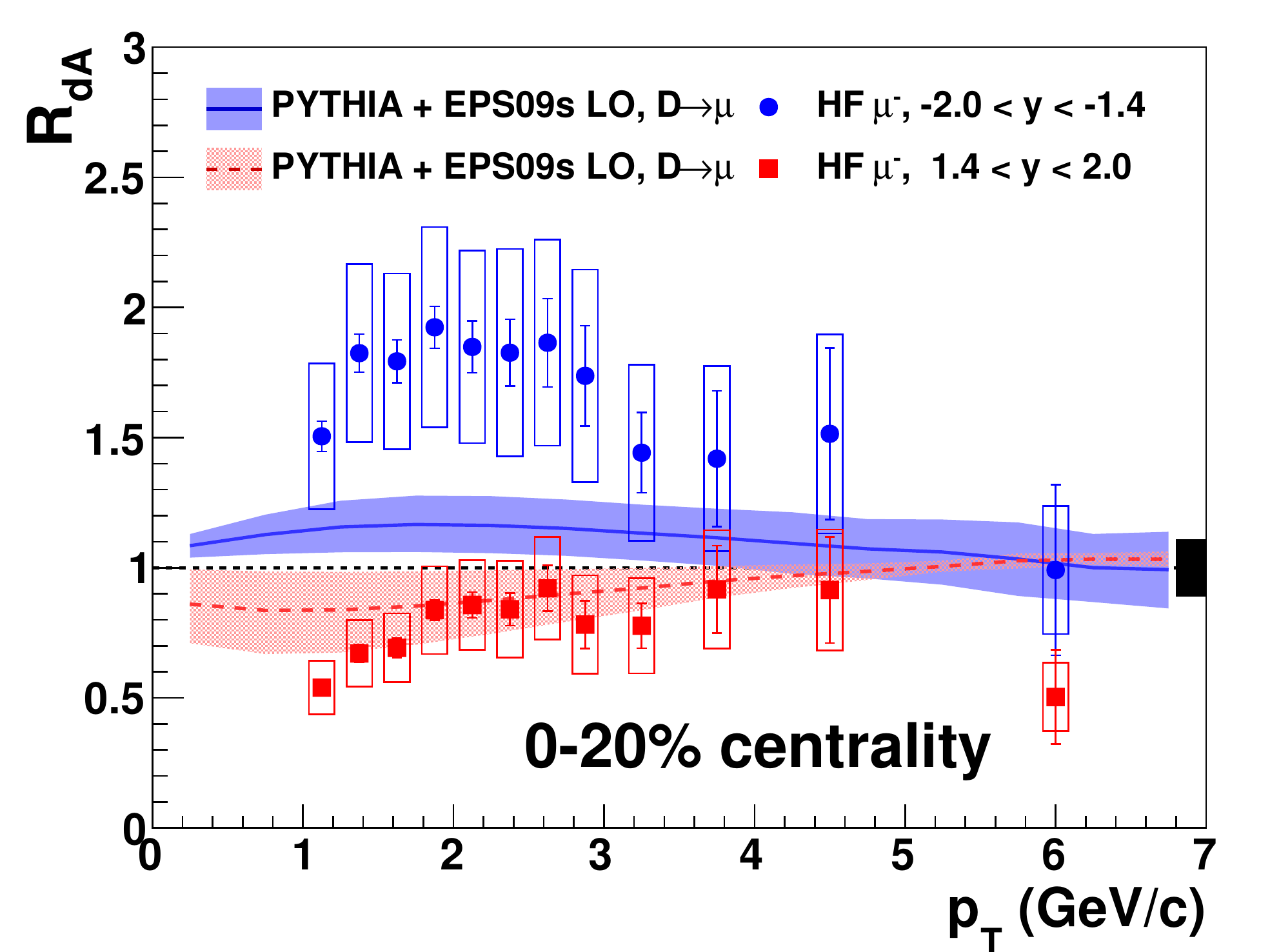}
\includegraphics[width=0.49\textwidth,clip]{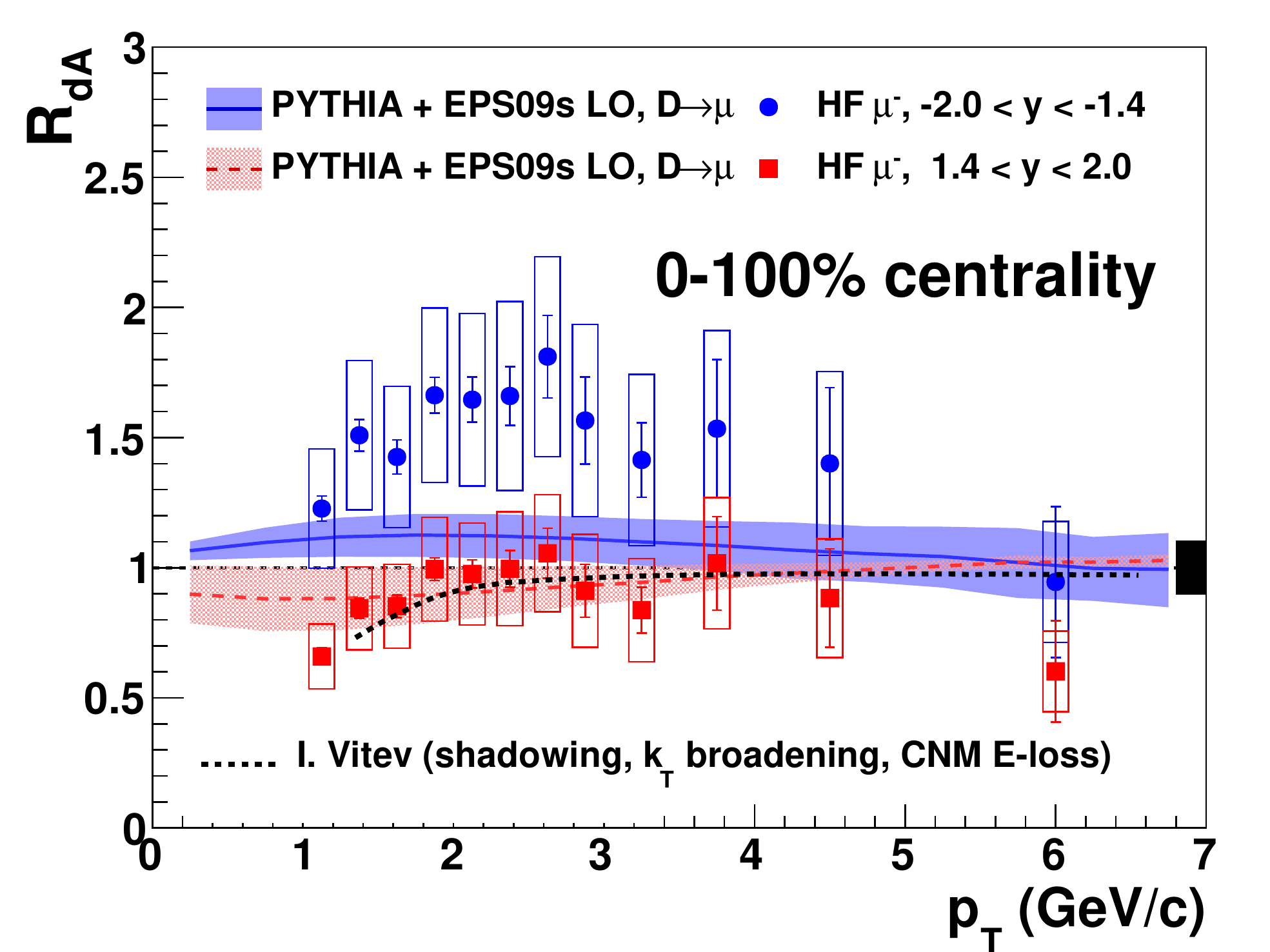}
\caption{\rda of heavy-flavor muons as a function of \pt at forward and backward rapidity regions in the most peripheral (top left), the most central (top right), and the unbiased (bottom) \dau collisions compared with two theoretical predictions, PYTHIA$+$EPS09s nPDF and pQCD calculations.}
\label{fig3}
\end{figure}

Recently, heavy-flavor muons have been measured for various \dau centrality classes at forward and backward rapidity regions~\cite{ppg153}.
Figure~\ref{fig3} shows \rda as a function of \pt for heavy-flavor muons at three centrality ranges, 60--88\% (top left), 0--20\% (top right), and 0--100\% (bottom), at forward (squares) and backward (circles) rapidity.
In the most peripheral centrality class, \rda at both rapidity regions are consistent with the unity.
However, a clear enhancement is observed at backward rapidity where Au nucleus (high $x$) is going in the most central collisions.
At forward rapidity which is $d$ direction (low $x$), heavy-flavor muons are supressed than the scaled \pp results.
From these comparisions, additional CNM effects beyond the nPDF modification may play an important role depending on rapidity.

Theoretical calculations, PYTHIA$+$EPS09s nPDF set~\cite{eps09s:2012}, considering only modification of nPDF are plotted in the same plots.
The prediction shows a qualitative agreement with the forward data, but it underestimates the enhancement seen in backward rapidity and the differece between forward and backward.
Another theoretical approach from pQCD calculation in the unbiased centrality class is consistent with the forward data.
This pQCD prediction considers the CNM effects such as shadowing, \pt broadening, and energy loss.

\section{Summary}
\label{sum}
PHENIX have measured leptons from open heavy-flavor decay in variety of collision systems.
Following points can summarize the obtained results.
\begin{itemize}
\item A nice trend between cold and hot nuclear matter effects depending on the system size (\npart) is observed; a clear enhancement in central \dau, a small suppression in central \cucu, and a significant suppression in central \auau collisions.
\item In \dau collisions, an enhancement is observed at mid- and backward rapidity regions, whereas a suppression is seen at forward rapidity. The prediction from the EPS09s nPDF model shows a similar trend of rapidity dependence, but it underestimate the difference between forward and backward rapidity seen in the data.  
\end{itemize}

\bibliographystyle{elsarticle-num}







\end{document}